# Combined narrowband imager-spectrograph with volume-phase holographic gratings

Eduard R. Muslimov*[a,b], Sergei N. Fabrika[c], Gennady G. Valyavin[c]
[a]Aix Marseille Univ, CNRS, LAM, Laboratoire d'Astrophysique de Marseille, Marseille, France;
[b]Kazan National Research Technical University named after A.N. Tupolev-KAI 10 K. Marx, Kazan, Russian Federation 420111, [c]Special Astrophysical Observatory of the Russian AS, Nizhnij Arkhyz, Russian Federation 369167

## ABSTRACT

In the present work we discuss a possibility to build an instrument with two operation modes - spectral and imaging ones. The key element of such instrument is a dispersive and filtering unit consisting of two narrowband volume-phase holographic gratings. Each of them provides high diffraction efficiency in a relatively narrow spectral range of a few tens of nanometers. Besides, the position of this working band is highly dependent on the angle of incidence. So we propose to use a couple of such gratings to implement the two operational modes. The gratings are mounted in a collimated beam one after another. In the spectroscopic mode the gratings are turned on such angle that the diffraction efficiency curves coincide, thus the beams diffracted on the first grating are diffracted twice on the second one and a high-dispersion spectrum in a narrow range is formed. If the collimating and camera lenses are corrected for a wide field it is possible to use a long slit and register the spectra from its different points separately. In the imaging mode the gratings are turned to such angle that the efficiency curves intersect in a very narrow wavelength range. So the beams diffracted on the first grating are filtered out by the second one except of the spectral component, which forms the image. In this case the instrument works without slit diaphragm on the entrance. We provide an example design to illustrate the proposed concept. This optical scheme works in the region around 656 nm with F/# of 6.3. In the spectroscopic mode it provides a spectrum for the region from 641 to 671 nm with reciprocal linear dispersion of 1.4 nm/mm and the spectral resolving power higher than 14000. In the imaging mode it covers linear 12mm$^{\text{X}}$12mm field of view with spatial resolution of 15-30 lines/mm.

## 1. INTRODUCTION

High-resolution spectroscopy in a selected waveband represents a powerful diagnostics tool, which is in use for different applications. A good example can be found in the field of nuclear fusion, where spectral measurements around specific emission lines like the carbon triplet[1] or the H-alpha line[2,3] are used to control the plasma state. On the other hand, in the same application field there are cases when the image in a narrow spectral range is registered[4,5]. In the first case a spectrograph with a complex dispersive unit[1] can be used, while in the second one interference filter represents a typical solution. In the present article we consider a possibility to combine both of the measurement modes in a single instrument.

Before one of the authors had considered an optical scheme of a high-resolution narrowband spectrograph with a pair of volume-phase holographic (VPH) gratings[6]. It was shown that in spite of high diffractive efficiency for the central wavelength the high-dispersion gratings are characterized by strong spectral selectivity. When two gratings are mounted one after another it causes huge transmission losses on the spectrum edges, which limits application of such a scheme in spectrographs. In addition, the scheme is very sensitive to the gratings positioning. However, in the context of the present problem these features can be turned into advantages. Spatial and angular selectivity are well-known properties of VPH gratings[7], which were used several times spectral and spatial filtering[8] and for creation of tunable filters[9].

We propose an optical scheme which also uses two VPH gratings to provide high dispersion in the spectral mode. To switch to the imaging mode both of the gratings are turned to such an angle that their individual diffraction efficiency

(DE) curves intersect each other in a narrow region. Implementation of this simple concept requires a special approach to the design process because numerous mutually connected target parameters should be controlled in two modes. The paper is organized as follows. In Section 2 we present optical scheme, exemplifying the above-mentioned idea and provide estimation of its image quality in two modes. Section 3 contains description of the instrument throughput and bandwidth control. In Section 4 we discuss the obtained results, limitations and possible improvements of this design.

## 2. OPTICAL DESIGN

To demonstrate the proposed concept we consider optical scheme of a combined instrument working around H-alpha line in two modes. The initial data taken for the design is summarized in Table 1. We should note that these values are not connected to any existing application or device and just approximately correspond to the currently available components and instruments.

Table 1. Initial data for the instrument design.

| **Wavelength, nm** | **Spectral mode** | **Imaging mode** |
|---|---|---|
| Wavelength, nm | 641-671 | 656 |
| Entrance diaphragm size, mm$^2$ | 12$^x$0.025 | 12$^x$12 |
| Image size, mm$^2$ | 12$^x$21.5 | 12$^x$12 |
| F/# | 6.3 | 6.3 |
| Linear magnification | 1 | 1 |
| | | |

### 2.1. Optical scheme

The optical scheme consists of two lenses – collimator and camera and two VPH gratings. The aperture and field of view are relatively moderate so it is enough to use simple triplet-type lenses. Focal length of each of them is 150 mm. The gratings operate with large angles of incidence and diffraction. The rotation angles sum is equal to 180º to keep simple and compact layout. Each of the gratings is recorded in a photosensitive material (e.g. dichromated gelatin – DCG) on a glass substrate and protected by a cover glass. The first grating has spatial frequency of 2284 mm$^{-1}$, and the second one has 1957 mm$^{-1}$. The general view of the scheme in two modes is pre

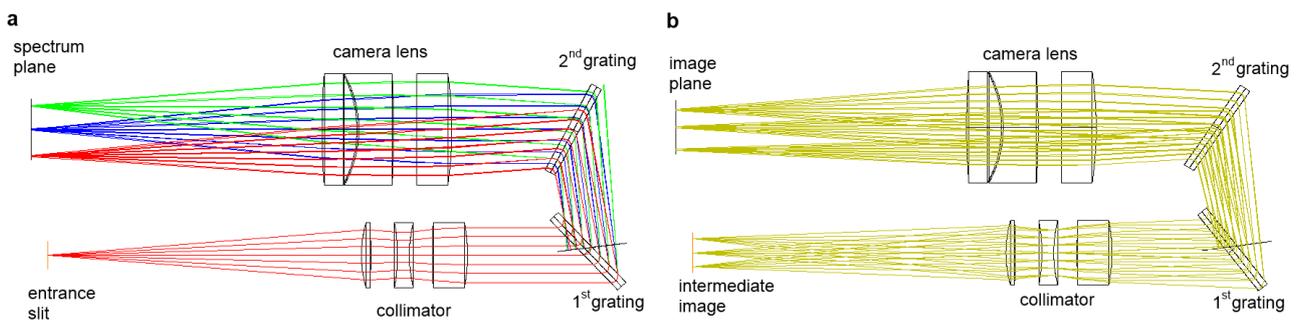

Figure 1. Optical scheme of the instrument: a – spectral mode, b – imaging mode.

### 2.2. Image quality

To estimate the optical quality achieved in the spectral mode the instrument functions (IF) are used. The IF calculated for the entrance slit width of 25 µm are plotted on Fig.2.

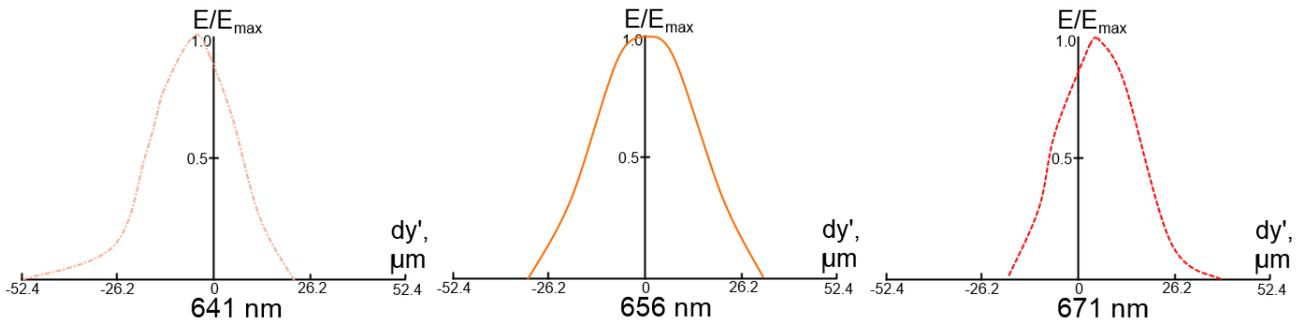

Figure 2. Instrument functions in the spectral mode for the entrance slit width of 25 μm.

Using the IF data and the reciprocal linear dispersion value, which equal to 1.395 nm/mm, it is possible to compute the spectral resolution and the unitless spectral resolving power (see Table 2). One can see that the spectral resolution is high over entire the working range. In addition we can note here that these numbers almost do not change for different X coordinate values, so measurements with a long slit are possible.

Table 2. Instrument resolution in the spectral mode.

| Wavelength, nm | Instrument function FWHM, μm | Spectral resolution, nm | Spectral resolving power |
|---|---|---|---|
| 641 | 26.4 | 0.037 | 17402 |
| 656 | 33.6 | 0.047 | 13993 |
| 671 | 25.6 | 0.036 | 18785 |

For estimation of the instrument spatial resolution in the imaging mode the spot diagrams and module transfer function (MTF) are used. The spot diagrams (Fig. 3a) has root-mean square of 17.3-19.9, 12.4-13.9 and 9.0-13.9 μm at 641, 656 and 671 nm, respectively. The spatial resolution defined by the MTF varies from 15 to 30 mm$^{-1}$.

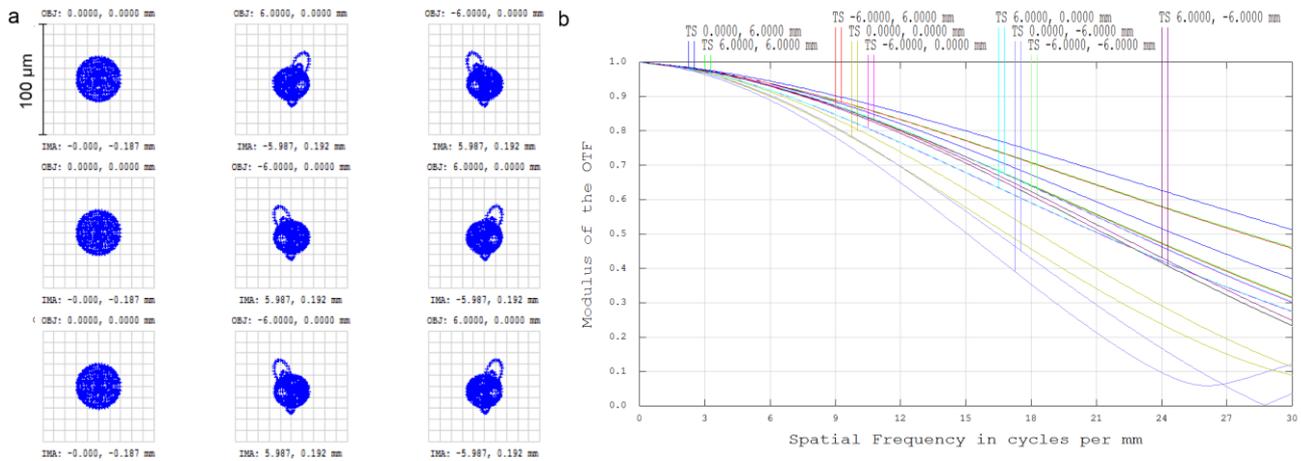

Figure 3. Image quality in the imaging mode: a – spot diagrams, b – module transfer function.

## 3. THROUGHPUT CONTROL

### 3.1. Diffraction efficiency modeling and optimization

If the frequency and angle of incidence of a VPH grating are given, its efficiency depends on the hologram thickness, index modulation and inclination of the fringes. However, angle of incidence on the second grating is determined by diffraction on the first one, so the DE curves of the gratings cannot be optimized separately. A script, which performs principal ray tracing and then computes DE of each of the gratings for a separate beam, was used for calculation and optimization of the gratings pair throughput. The DE computation was performed with use of rigorous coupled-wave analysis (RCWA) method implemented in *reticolo* software[10,11].

Then the gratings throughput can be optimized by means of ordinary procedure[12]. The difference in this case consists of control on the angles of incidence. For the spectral mode they are defined on the previous design stage and cannot be changed, while for the imaging mode they represent free optimization parameters, which are used to achieve desired centering and bandwidth of the throughput curve. In the latter case special boundary condition defining sum of the rotation angles should be applied. The gratings parameters obtained after the optimization are listed in Table 3.

Table 3. VPH gratings parameters.

| **Parameter** | **1st grating** | **2nd grating** |
|---|---|---|
| **Fringes frequency, mm$^{-1}$** | 2284 | 1957 |
| **Holographic layer thickness, µm** | 12 | 6.3 |
| **Index modulation** | 0.027 | 0.048 |
| **Fringes inclination** | 2.15 | -6.9 |
| **Rotation angle in the spectral mode, º** | 43.94 | 46,5 |
| **Rotation angle in the imaging mode, º** | 54.87 | 48.9 |

The curves of DE for individual grating and the total throughput for each of the modes are shown on Fig.4.

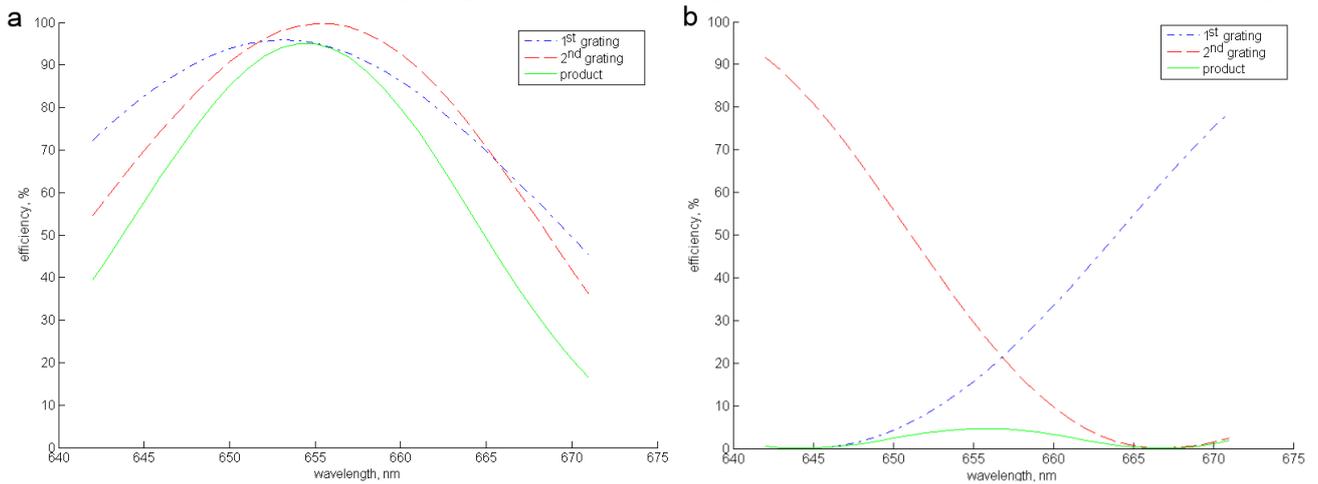

Figure 4. Diffraction efficiency curves: a – spectral mode, b – imaging mode.

The graphs clearly show that the total bandwidth becomes narrower with rotation of the gratings. However, the peak transmission value significantly decreases. It would be possible to minimize this undesired effect if the DE curves of the gratings had sharp edges. Such a property can be achieved for one edge of VPH grating efficiency curve[13], but it cannot be repeated with an arbitrary working geometry. In the present case it was found to be impossible to meet this additional condition with a limited number of free parameters. Other means like change of the fringes profile[14] do not have any significant effect.

On the other hand, the throughput parameters change continuously with rotation of the gratings, thus allowing to tune the instrument performance. Values for intermediate positions of the gratings are given in Table 4.

Table 4. Change of the total throughput.

| 1st grating rotation angle, º | 2nd grating rotation angle, º | Throughput curve FWHM, nm | Peak transmission, % | Spectral image length, mm |
|---|---|---|---|---|
| 43.94 | 54.87 | 12.0 | 94.92 | 21.50 |
| 44.5 | 53.7 | 11.4 | 89.87 | 21.74 |
| 45.0 | 52.5 | 9.7 | 67.86 | 22.03 |
| 45.5 | 51.3 | 8.0 | 39.93 | 22.34 |
| 46.0 | 50.1 | 5.5 | 17.07 | 22.66 |
| 46.5 | 48.9 | 5.2 | 4.81 | 23.00 |

### 3.2. Images simulation

To demonstrate performance of the developed scheme we simulated images in the both of working modes. They were produced by means of non-sequential ray tracing, which allows to account for actual diffraction efficiency and rays splitting. In each case 7 wavelengths covering the range from 642 to 670 nm were traced. The object was presented by an array of point sources. For considerations of simplicity each of them had power of 1W. For the spectral mode 3 sources with coordinates equal to (-6 mm, 0 mm), (0 mm, 0 mm) and (6 mm, 0 mm) were used. The of ray tracing are shown on Fig.5 (the image is slightly defocused to provide better seeing). Positive direction of the Y coordinate on the graph corresponds to increase of the wavelength.

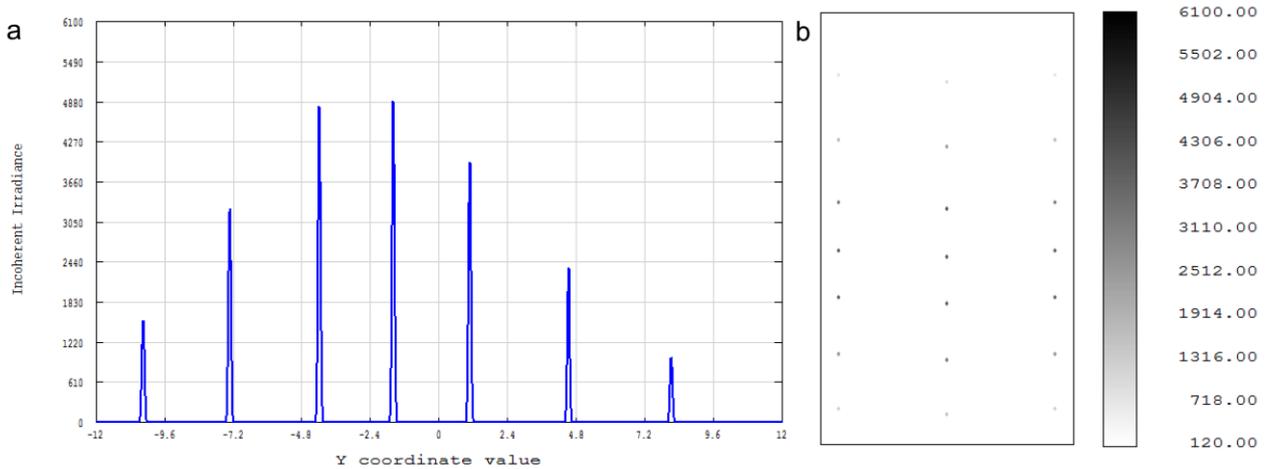

Figure 5. Simulated image in the spectral mode: a – column cross-section, b – gray-scale image (the irradiation units are W/cm$^2$).

The same simulation was repeated with the imaging mode configuration. This time 9 point sources with X and Y coordinates equal to -6, 0 or 6 mm were used. The results are presented on Fig.6. The image corresponding to wavelength of 656 nm is highlighted. Except of the target wavelength there are other spectral components present in the image. Their influence can be decreased, if the detector has a cut-off level (like the dashed line shown on Fig.6a). Such a cut-off is accounted for on Fig.5b and 6b. One can see that it has a negligible influence on the spectral image. But allows to separate the desired band in the narrowband image. Besides, the residuals in the image can be excluded if spectrum of the observed object is known in advance.

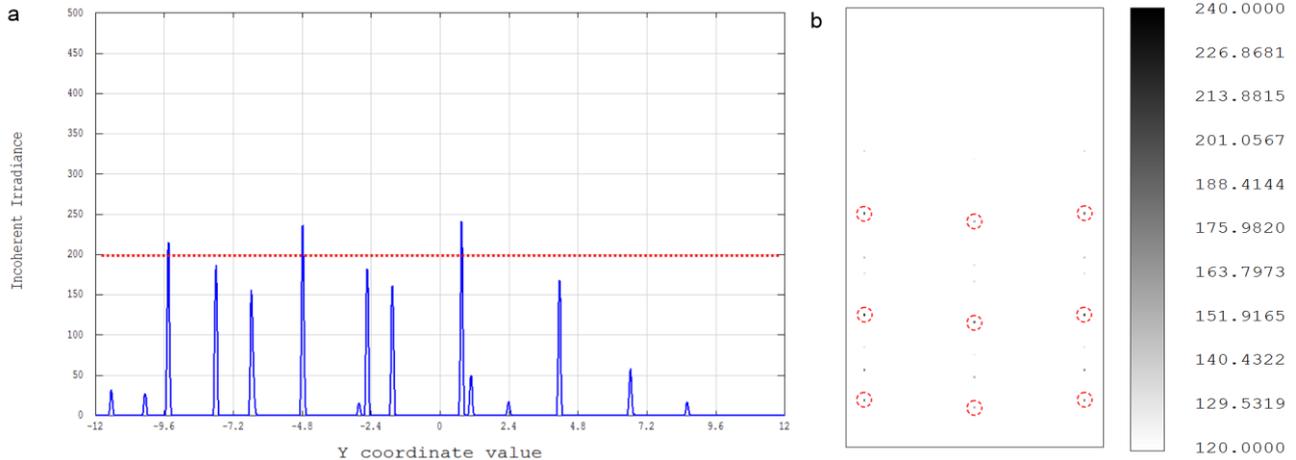

Figure 6. Simulated image in the imaging mode: a – column cross-section, b – gray-scale image (the irradiation units are W/cm$^2$).

This scheme may be really very useful for astrophysics in a search for massive stars like LBV (Luminous Blue Variables), and very massive stars (VMS) with masses exceeding 150 solar masses, and X-rays sources in nearby galaxies. Using imaging mode one may cover the whole galaxy to detect emission objects (H-alpha line) to separate HII regions which are very numerous in galaxies (the line width less than 20 km/sec) and to detect LBVs, whoes line width bigger than 150-300 km/sec (spectral resolution higher than 1000-2000). After selection the potential LBV stars one may change to spectral mode to find the massive stars with strong winds. There are no many LBVs in nearby galaxies [15-17], however discovery such objects is very important. The same is valid for VMSs, where only a few such stars were found[18]. It is very essential to discover new SS433-like objects in galaxies[19-21], they are black holes with super-Eddington accretions disks because only one such object (SS433) has been found in the Milky Way. An absolutely new topic is detection of Ultraluminous X-ray Sources (ULXs)[22,23], no one ULX has been found in the Milky Way and even in the Local Group of galaxies. However, up to 50 Mpc in the local Universe about 30 ULXs have been detected, and only 10 such objects have been detected with optical spectra. The ULX-search is very significant; as a rule they are discovered in galaxies with a strong star formation rate.

## 4. CONCLUSION

In the present paper we considered optical design of a tunable narrowband spectrograph. It operates around H-alpha with high dispersion. The dispersion is provided by two VPH gratings mounted one after another in collimated beam. By rotation of these gratings it is possible to control the instrument bandwidth, which changes by factor of 2.3. When the gratings are turned by 3.56º and 5.97º respectively. So in the extreme positions two modes are implemented. In the first one a bright high-dispersion spectrum for a long slit is formed. The spectral resolving power in the range of 641-671 nm is 13993-18785. In the second one the instrument creates a narrowband image for a square field of view. Spatial resolution in this imaging mode is 15-30 mm$^{-1}$. The optical scheme is notable for its simplicity and compactness.

Numerical simulations demonstrate that in the spectral mode the instrument provides high irradiation of the image. In the imaging mode there is a visible loss of energy and a risk of dispersive blurring of the image. These problems are caused by inappropriate shape of the diffraction efficiency curves. They can be partially solved by control of the detector cut-off or use of *a priori* information about the object.

The proposed scheme can be of special interest for some diagnostics applications like that in the field of nuclear fusion. However, its possible application can be significantly extended on account for astrophysics.

## ACKNOWLEDGEMENT

This work was supported by the Russian Science Foundation (project No. 14-50-00043). Eduard Muslimov acknowledges the support of his personal post-doc contract by the European Research council through the H2020 - ERC-STG-2015 – 678777 ICARUS program. The authors also thank J.-P. Hugonin and Ph. Lalanne from Institut d'Optique for provision of the *reticolo* software used in this work.